\documentclass[conference]{IEEEtran}
\IEEEoverridecommandlockouts

\usepackage{cite}
\usepackage{amsmath,amssymb,amsfonts,amsthm}
\usepackage{mathtools}
\usepackage{algorithm}
\usepackage{algorithmic}
\usepackage{graphicx}
\usepackage{textcomp}
\usepackage{xcolor}
\usepackage{gensymb}
\usepackage{verbatim}
\def\BibTeX{{\rm B\kern-.05em{\sc i\kern-.025em b}\kern-.08em T\kern-.1667em\lower.7ex\hbox{E}\kern-.125emX}}

\DeclarePairedDelimiter{\round}\lfloor\rceil

\newtheorem{proposition}{Proposition}

\newtheorem{lemma}{Lemma}
\newtheorem{remark}{Remark}

\newtheorem{corollary}{Corollary}

\newcommand{\rmd}{{\,\mathrm{d}}}
\renewcommand{\IEEEQED}{\IEEEQEDopen}

\begin{document}
\bstctlcite{IEEEexample:BSTcontrol}

\title{Performance Analysis of RIS-Assisted Large-Scale Wireless Networks Using Stochastic Geometry}

\author{
        Tianxiong Wang,
        Gaojie Chen,~\IEEEmembership{Senior Member, IEEE,}
        Mihai-Alin~Badiu,
        \\ and Justin~P.~Coon,~\IEEEmembership{Senior Member, IEEE,}
\thanks{T. Wang, M. A. Badiu and J. P. Coon are with the Department of Engineering Science, University of Oxford, Oxford, OX1 3PJ, U.K. (e-mail: \{tianxiong.wang, mihai.badiu, justin.coon\}@eng.ox.ac.uk).}
\thanks{G. Chen is with 5GIC \& 6GIC, Institute for Communication Systems (ICS), University of Surrey, Guildford, GU2 7XH, United Kingdom (e-mail: gaojie.chen@surrey.ac.uk).}
\vspace{-2em}}

\maketitle

\begin{abstract}
    In this paper, we investigate the performance of a reconfigurable intelligent surface (RIS) assisted large-scale network by characterizing the coverage probability and the average achievable rate using stochastic geometry. Considering the spatial correlation between transmitters (TXs) and RISs, their locations are jointly modelled by a Gauss-Poisson process (GPP). Two association strategies, i.e., nearest association and fixed association, are both discussed. For the RIS-aided transmission, the signal power distribution with a direct link is approximated by a gamma random variable using a moment matching method, and the Laplace transform of the aggregate interference power is derived in closed form. Based on these expressions, we analyze the channel hardening effect in the RIS-assisted transmission, the coverage probability, and the average achievable rate of the typical user. We derive the coverage probability expressions for the fixed association strategy and the nearest association strategy in an interference-limited scenario in closed form. Numerical results are provided to validate the analysis and illustrate the effectiveness of RIS-assisted transmission with passive beamforming in improving the system performance. Furthermore, it is also unveiled that the system performance is independent of the density of TXs with the nearest association strategy in the interference-limited scenario.
\end{abstract}


\section{Introduction}
Wireless connectivity and mobile data traffic have increased remarkably in the recent years \cite{jiang2021road}. As foreseen by \cite{dang2020should}, this trend will continue in the future sixth-generation (6G) wireless communication networks, which are expected to provide highly efficient and reliable wireless communication services for billions of devices. The significant demands in high data rates, reliability and massive connectivity have brought great challenges in designing wireless systems and spurred enthusiasm in developing innovative physical layer technologies. Among various potential technologies, reconfigurable intelligent surfaces (RISs) have recently been regarded as a key enabling solution owing to the theoretic and engineering development of metasurfaces \cite{tang2020wireless}. Physically, RIS is a planar metasurface equipped with a smart RIS controller and a large number of passive and low-cost elements \cite{Q2020T}. Each reflecting element can be controlled independently by a RIS controller to adjust the phase and amplitude of the reflected signals intelligently, hence collaboratively altering the wireless propagation environment for achieving various desired purposes, such as directional signal enhancement and spatial signal nulling \cite{wu2021intelligent}. Compared to some related technologies, RIS has advantages in energy efficiency, low hardware cost, and convenient deployment \cite{b5}. Hence, the application of RIS is extensively investigated in different areas, such as RIS empowered physical layer security \cite{X.P21,H.C21}, RIS empowered non-orthogonal multiple access (NOMA) \cite{pan2021reconfigurable}, RIS empowered unmanned aerial vehicle (UAV) communications \cite{yang2020performance,F.S21}, etc. Besides, RIS is expected to be widely deployed in future wireless communication networks to improve communication quality and extend coverage.  

Due to the attractive benefits of RIS, extensive research efforts have been devoted to the performance analysis of RIS-assisted communication systems to analyze the link-level performance gains that can be brought by integrating RISs into wireless networks. The authors in \cite{ozdogan2019intelligent} derived the path loss expressions of RIS-assisted wireless networks and illustrated that the path loss is inversely proportional to the product of the distances of two cascaded links. Characterizing the small-scale fading, the authors in \cite{bb1} derived the approximate outage probability, symbol error probability and average achievable rate of a RIS-assisted system over Rayleigh fading channels without phase errors by applying the central limit theorem (CLT). As a step further, the phase errors of RIS were considered in \cite{b8} and the authors proved that the composite channel of a RIS system with phase errors is Nakagami-$m$ fading. However, as shown in \cite{9681955, b9}, the approximation methods based on the CLT are only accurate for a very large number of elements, and the approximation errors are significant at the high transmit signal-to-noise ratio (SNR) regime. The authors in \cite{cui2021snr} proposed a moment matching approach to approximate the composite channel gain as a gamma random variable and then derive various performance metrics in closed form. The moment matching method turns out to be accurate even for a small number of elements. Following a similar method, the outage probability of a RIS-aided network with 1-bit phase shifting was derived in \cite{9322575}. All the above works considered the Rayleigh fading channel model. To further investigating the general channel environment, the Nakagami-$m$ fading model was applied in \cite{ibrahim2021exact, tegos2021distribution}. For example, the authors in \cite{ibrahim2021exact} derived the outage probability of a RIS aided network without phase errors over Nakagami-$m$ channels in an integral form by using the Gil-Pelaez theorem. Under the random phase-shifting scheme where the phase errors are uniformly distributed over $(0, 2\pi)$,  the authors in \cite{tegos2021distribution} derived the outage probability in closed form by assuming Nakagami-$m$ fading channels. Recently, the effects of channel correlations were investigated in \cite{van2021outage} and \cite{9774334}, where the outage probability of a RIS-assisted wireless network over correlated channels with random and coherent phase-shifting schemes were derived. 

The aforementioned works all consider single-cell networks. However, a real RIS-assisted communication network comprises multiple transmitters, receivers and RISs, and they may share the same resource block, i.e., time and frequency. Thus, in the performance analysis of RIS-assisted communication systems, it is essential to model the RIS-assisted multi-cell network subjected to inter-cell interference. Stochastic geometry has been widely used to model and characterize the performance of large-scale wireless networks, which can provide useful performance bounds and practical insights regarding the practical wireless systems \cite{haenggi2012stochastic, C.G17}. Performance analysis of RIS-assisted multi-cell networks using stochastic geometry has been carried out in recent works. The authors in \cite{kishk2020exploiting} modelled the blockages as boolean line segments, and the RISs are attached to the blockages with a probability. The probability that a given user is associated with a transmitter (TX) via a RIS is derived, which shows that RISs can enlarge the coverage in a multi-cell network by providing indirect LoS paths. In \cite{zhu2020stochastic}, the authors modelled the locations of TXs and RISs as two independent homogeneous Poisson point processes (HPPP) in a multi-cell millimetre-wave (mmWave) network and derived the average achievable rate in a multi-fold integral. With a similar system setup, the authors in \cite{lyu2021hybrid} analyzed the distributions of the signal power and interference power, respectively and derived a two-fold integral expression for the coverage probability of the typical user when it is connected with the nearest TX and RIS. RIS-assisted multi-cell NOMA networks were investigated in \cite{9606895}, in which the authors modelled the locations of TXs and users as HPPP and assumed the interference from other RISs that are not associated with the typical user could be ignored. The coverage probability and average achievable rate of the typical user were both derived in integral form. The performance of two different types of users, i.e., served by the TX with and without RIS-assisted transmissions, are analyzed in a more recent work \cite{9673721}, where the authors derived the coverage probability of the network conditioned on the proportions of the two types of users. 

All the aforementioned works use independent HPPP to model the locations of TXs and RISs, because HPPP not only can model the randomly located nodes but also makes the analysis mathematically tractable \cite{haenggi2012stochastic}. However, the HPPP may not be an accurate model for large-scale RIS-assisted systems since it does not capture the spatial correlations between TXs and RISs \cite{guo2016gauss}.  Since RIS is a passive device and its locations have to be appropriately designed according to the location of the TX, the locations of TXs and RISs exhibit significant spatial correlations \cite{Q2020T}. To describe this characteristic of the large-scale RIS-assisted networks, the Gauss-Poisson process (GPP) can be applied because the GPP is a clustering process that can capture the correlations of the node locations, in which each cluster has one or two points with the spatial correlations \cite{deng2018benefits}. 

Therefore, we use a GPP to model a large-scale RIS-assisted network to represent the spatial correlations between TXs and RISs. Furthermore, the general Nakagami-$m$ fading channel model and the RIS-assisted transmission with the direct link between the user and the TX are considered in the analysis. The main contributions of this paper are listed as follows:
\begin{enumerate}
    \item To provide an accurate system-level performance analysis, we propose an analytical framework to characterize the performance of a RIS-assisted large-scale network by using stochastic geometry. The locations of UEs are modelled by an HPPP, while the locations of TXs and RISs are modelled by a cluster process GPP to characterize the spatial correlations between TXs and RISs. The channels related to the RIS are Nakagami-$m$ fading. Moreover, we consider the fixed association and nearest association strategies between the TX and the typical user in the analysis, respectively.
    \item For the RIS-aided downlink transmission with a direct link, the desired signal power is approximated by a gamma random variable using a moment matching method. Furthermore, the distribution of the interference power scattered by a RIS is approximated by the exponential distribution. Then, we derive the closed form Laplace transform of the aggregate interference power from the interfering TXs and RISs for both association strategies. 
    \item We obtain expressions for the coverage probability and the average achievable rate in the downlink for the typical user with the fixed and nearest association strategies. The expressions for the coverage probability have the closed forms for both association strategies in terms of elementary and/or special functions for all path loss exponents. The expressions for the average achievable rate have integral forms, which are further simplified to closed forms for certain path loss exponents.
    \item The numerical results verify the effectiveness of the theoretical analysis, and practical insights are extracted from the analytical and numerical results. For the fixed association strategy, increasing the density of TXs and the RIS association probability will both impair the system performance due to the enhanced interference power. While for the nearest association strategy, a larger density of TXs and RIS association probability is beneficial since the typical user is more likely to be served by a closer TX with an assisting RIS. Furthermore, we prove that the coverage probability and the average achievable rate are unrelated to the density of TXs when the nearest association strategy is applied in the large transmit SNR regime.
\end{enumerate}

The rest of the paper is organized as follows. The system model is introduced in Section II. The distributions of the desired signal power and the interference power for two types of association strategies are analyzed in Section III and Section IV, respectively. In Section V, the coverage probability and average achievable rate are analyzed. Section VI gives the numerical results, and the paper is concluded in Section VII.

\textit{Notations}: The expectation and variance of a random variable $X$ are denoted as $\mathbb{E}[X]$ and ${\rm Var}[X]$, respectively. The cumulative distribution function (CDF) and the complementary CDF (CCDF) of a random variable $X$ are denoted as $F_{X}(x)$ and $\bar{F}_{X}(x)$, respectively. $\mathbb{C}$ and $\mathbb{C}^{N \times 1}$ represent the complex numbers and the $N \times 1$ complex vectors space, respectively. $\mathcal{CN}(\mu, \sigma^2)$ denotes the complex Gaussian distribution with mean $\mu$ and variance $\sigma^2$. $\mathcal{R}(\sigma)$ denotes the Rayleigh distribution with the scale parameter $\sigma$. 
$\Gamma(\cdot)$, $\Gamma(\cdot, \cdot)$ and $\gamma(\cdot, \cdot)$ stand for the gamma function, upper incomplete gamma function and lower incomplete gamma function, respectively. $\round{\cdot}$ is the rounding operation. ${_{2}F_1}\left(\cdot, \cdot; \cdot; \cdot\right)$ is the hypergeometric function and $\sim$ is the notation for the asymptotic equivalence. ${\rm erfc(\cdot)}$, ${\rm Ci(\cdot)}$ and ${\rm Si(\cdot)}$ denote the complementary error, cosine integral and sine integral functions, respectively.

\section{System Model}
\begin{figure}[t!]
    \centerline{\includegraphics[scale=0.15]{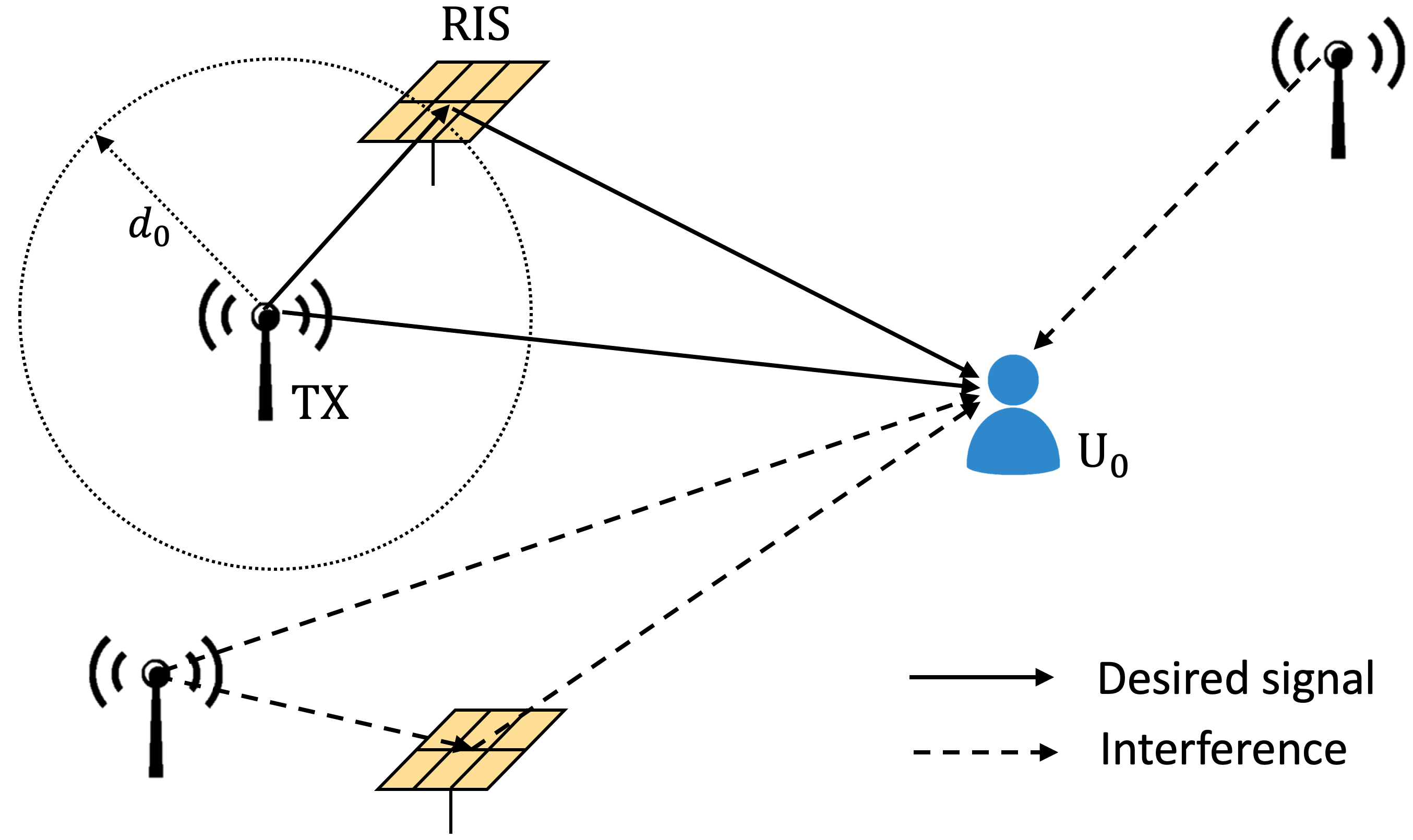}}   
    \caption{System model of RIS assisted large-scale network.}
    \label{sys1}
\end{figure}

This paper studies a RIS-assisted multi-cell wireless communication network with multiple single-antenna transmitters (TXs), RISs, and single-antenna user equipment (UEs)\footnote{Analysis of multi-antenna TXs and UEs will be left for our future work. Moreover, as shown in \cite{lyu2021hybrid}, RIS-assisted communications with single-antenna TXs can achieve significant performance gain, especially in scenarios where multi-antenna systems cannot be equipped.} shown in Fig.~\ref{sys1}. The performance of the downlink communication from the TXs to UEs is analyzed. The locations of UEs are modelled by a homogeneous Poisson point process (HPPP) $\Lambda_u$ with density $\lambda_u$. We consider the spatial correlations between the locations of TXs and RISs, and model the locations of TXs and RISs as a GPP $\Lambda_c$. $\Lambda_c$ is a Poisson cluster process where the locations of TXs are the parent process and follow a HPPP $\Lambda_t$ with density $\lambda_t$. Each cluster in $\Lambda_c$ has one or two points. If a cluster has one point, it is at the location of the parent point. Otherwise, one of them is the parent point, and the daughter point, which represents the location of an assisting RIS of the corresponding TX, is uniformly distributed on a circle with a radius $d_0$ centred at the parent point\footnote{In this paper, we consider the standard GPP where $d_0$ is fixed for clarity. The proposed results can be easily extended to the generalized GPP model by averaging the derived coverage probability and average achievable rate over the distribution of $d_0$.}. The probabilities of each cluster having one or two points are $1 - p$ and $p$, respectively. In other words, for a typical TX, the probability that it has an assisting RIS is $p$. For notational simplicity, we denote the cluster processes with one point and two points as $\Lambda_1$ and $\Lambda_2$, respectively. Based on the definition of GPP, we have
\begin{equation}
    \Lambda_c = \Lambda_1 \cup \Lambda_2. 
\end{equation}
Furthermore, the parent and daughter point processes of the clusters with two points are respectively denoted as $\Lambda_2^{(P)}$ and $\Lambda_2^{(D)}$ such that 
\begin{equation}
    \Lambda_2 = \Lambda_2^{(P)} \cup \Lambda_2^{(D)}. 
\end{equation}

Since the locations of UEs are assumed to follow a HPPP, we can randomly select a typical UE $U_0$ from $\Lambda_u$ and fix it at the origin of the considered plane. As per Slivnyak's theorem, the performance of $U_0$ represents the average UE performance\footnote{In practice, each TX does not necessarily serve only one UE. As shown in \cite{zheng2020intelligent}, to fully exploit the performance gain of RIS, UEs associated with the same TX should be allocated with different time slots. In this paper, all the UEs are assumed to operate in the same time slot and frequency band.}. 
\subsection{Channel Model} 
It is assumed that each RIS has $N$ elements and can perform continuous phase adjustment to increase the composite channel gain of its serving UE\footnote{The assumption on the ideal coherent phase adjustment gives an upper bound on the system performance \cite{wang2021outage}. Furthermore, as proved in \cite{wu2019beamforming,wang2021performance}, only two-bit phase shifting leads to the performance close to upper bound in practice.}. The transmit power of each TX is $P$. The normalized small scale fading of the $T_i$-$U_0$ channel is denoted as $g_i \in \mathbb{C}$ where $i \in \Lambda_t$. The direct links between TXs and UEs are assumed to be Rayleigh fading channels, i.e., $|g_i| \sim \mathcal{R}(\sqrt{2}/2)$. The path loss of the direct link $T_i$-$U_0$ can be expressed as 
\begin{equation}
    \eta_{g, i} = C_d d_{g, i}^{-\alpha},
    \label{eta_g_i}
\end{equation}
where $d_{g, i}$ is the distance of the $T_i$-$U_0$ channel\footnote{Similar to \cite{xie2021star, hou2021mimo, xie2021modeling}, it is assumed that the horizontal distances of different channels are much larger than the vertical distances and the heights of TXs, RISs and UEs are ignored.}; $C_d$ is the path loss per unit distance of the direct link; $\alpha$ is the path loss exponent.

It is assumed that each RIS receives the signal from its associated TX and the signals from other TXs impinging on the RIS can be ignored\footnote{In this paper, we consider the spatial correlation between the locations of TXs and RISs. This assumption is particularly valid when the distances among different TXs are significantly larger than the distance between a TX and its associated RIS. A similar assumption is also applied in \cite{9606895,9673721}.}. For $T_j$, $j \in \Lambda_2^{(P)}$ that has an assisting RIS $R_{j'}$, $j' \in \Lambda_2^{(D)}$, the normalized small scale fading of the $T_j$-$R_{j'}$ and $R_{j'}$-$U_0$ channels are denoted as $\mathbf{h}_{j} = \left(h_{j, 1}, ... , h_{j, N}\right)^{T} \in \mathbb{C}^{N\times 1}$, and $\mathbf{r}_{j} = \left(r_{j, 1}, ... , r_{j, N}\right)^{T} \in \mathbb{C}^{N\times 1}$, respectively. $T_j$ and $R_{j'}$ are the parent and daughter points in the same two-point cluster. RISs operate in the far field of both TXs and UEs. We assume the reflected links associated with RISs are Nakagami-$m$ fading channels. To be specific, $|h_{j, n}| \sim {\rm Nakagami}(m_h, 1)$ and $|r_{j, n}| \sim {\rm Nakagami}(m_r, 1)$, where $n \in \{1, ..., N\}$. $m_h$ and $m_r$ are the shape parameters of $|h_{j, n}|$ and $|r_{j, n}|$, respectively. It is worth noting that the Nakagami-$m$ is a general distribution to model the channel fading and the Nakagami-$m$ distribution is equivalent to the Rayleigh distribution if the shape parameter equals to 1. The path loss of the reflected path $T_j$-$R_{j'}$-$U_0$ can be expressed as 
\begin{equation}
    \eta_{h, j} = C_r \left(d_{0} d_{r, j}\right)^{-\alpha},
    \label{eta_h_j}
\end{equation}
where $d_{r, j}$ is the distance of the $R_{j'}$-$U_0$ channel; $C_r$ is the path loss per unit distance of the reflected link.

\subsection{Signal Model}
The received signal at $U_0$ depends on whether its associated TX $T_0$ has an assisting RIS or not. If $T_0$ has an assisting RIS, the received signal at $U_0$ can be expressed as
\begin{equation}
    y_0^{+} = \left(\sqrt{\eta_{g, 0}} g_0 + \sqrt{\eta_{h, 0}} \mathbf{h}_0^{T} \mathbf{\Phi}_0 \mathbf{r}_0 \right) \sqrt{P} s_0 + \Tilde{I} + n_0,
    \label{re_sig_+}
\end{equation}
where
\begin{align}
    \Tilde{I} =& \sum\limits_{k \in \Lambda_t \backslash \{0\}} (1 - b_k) \sqrt{\eta_{g, k} P} g_k s_k \nonumber\\
    &+ b_k \left(\sqrt{\eta_{g, k}} g_k + \sqrt{\eta_{h, k}} \mathbf{h}_k^{T} \mathbf{\Phi}_k \mathbf{r}_k \right) \sqrt{P} s_k,
    \label{re_int}
\end{align}
representing the interference from all other TXs. $\mathbf{\Phi}_k = {\rm diag}\left(e^{j \phi_{k, 1}},..., e^{j \phi_{k, N}}\right)$ is the phase shifting matrix of $R_{k'}$, and $R_{k'}$ is the assisting RIS of $T_k$ in a two-point cluster; $b_k \in \{0, 1\}$ represents if $T_k$ is without or with an assisting RIS, respectively; $s_k$ is the unit power transmit symbol from $T_k$, satisfying $\mathbb{E}[s_k^2] = 1$ and $\mathbb{E}[s_k s_l] = 0$ for $k \neq l$; $n_0 \sim \mathcal{CN}(0, \sigma_0^2)$ is the additive white Gaussian noise (AWGN) with zero mean and variance $\sigma_0^2$. If $T_0$ does not have an assisting RIS, the received signal at $U_0$ can be expressed as
\begin{equation}
    y_0^{*} = \sqrt{\eta_{g, 0} P} g_0 s_0 + \Tilde{I} + n_0.
    \label{re_sig_*}
\end{equation}

If $T_0$ has an assisting RIS, the phase shifts of $R_0$ should be adjusted to align the phases of the reflected paths to the phase of the direct link, i.e.,
\begin{equation}
    \phi_{0, n} = -\arg(h_{0, n}) - \arg(r_{0, n}) + \arg(g_{0}).
\end{equation}
On the other hand, the phases of the composite channels $T_l$-$R_{l'}$-$U_0$, $l \in \Lambda_2^{(P)} \backslash \{0\}$ can be regarded as uniformly distributed over $[-\pi, \pi)$. Following the above discussions, \eqref{re_sig_+} can be written as
\begin{align}
    y_0^{+} = & \left(\sqrt{\eta_{g, 0}} |g_0| + \sqrt{\eta_{h, 0}} \sum\limits_{n = 1}^{N} |h_{0, n}| |r_{0, n}| \right) e^{j\arg(g_0)} \sqrt{P} s_0 \nonumber\\[5pt]
    & + \sqrt{P} \Tilde{I} + n_0,
    \label{re_sig_+1}
\end{align}
where
\begin{subequations}
\begin{align}
    \Tilde{I} &= \sum\limits_{k \in \Lambda_t \backslash \{0\}} b_k \Tilde{I}_{1, k} + (1 - b_k) \Tilde{I}_{2, k};\\
    \Tilde{I}_{1, k} &= \left(\sqrt{\eta_{g, k}} g_k + \sqrt{\eta_{h, k}} \sum\limits_{n = 1}^{N} |h_{k, n}| |r_{k, n}| e^{j \theta_{k,n}} \right) s_k;\\
    \Tilde{I}_{2, k} &= \sqrt{\eta_{g, k}} g_k s_k,
\end{align}
\end{subequations}
and $\theta_{k,n}$ is uniformly distributed over $[-\pi, \pi)$.

Then, we will give the SINR of $U_0$ at different association strategies as follows:
\begin{enumerate}
    \item With the fixed association strategy, the SINR of $U_0$ with an assisting RIS is given by
    \begin{equation}
        \gamma_0^{+} = \frac{S_0^{+}}{I + \gamma_t^{-1}},
    \end{equation}
    where $\gamma_t = P/\sigma_0^2$. The signal power $S_0^{+}$ is
    \begin{equation}
        S_0^{+} = \left(\sqrt{\eta_{g, 0}} |g_0| + \sqrt{\eta_{h, 0}} \sum\limits_{n = 1}^{N} |h_{0, n}| |r_{0, n}| \right)^2.
        \label{s0+}
    \end{equation}
    The interference power is
    \begin{equation}
        I = |\Tilde{I}|^2 = \sum\limits_{k \in \Lambda_t \backslash \{0\}} b_k I_{1, k} + (1 - b_k) I_{2, k},
        \label{i_power}
    \end{equation}
    in which
    \begin{subequations}
    \begin{align}
        I_{1, k} &= |\Tilde{I}_{1, k}|^2 \label{i_1kpower}\\
        &= \left|\sqrt{\eta_{g, k}} g_k + \sqrt{\eta_{h, k}} \sum\limits_{n = 1}^{N} |h_{k, n}| |r_{k, n}| e^{j \theta_{k,n}} \right|^2, \nonumber\\
        I_{2, k} &= |\Tilde{I}_{2, k}|^2 = \left|\sqrt{\eta_{g, k}} g_k\right|^2.\label{i_2kpower}
    \end{align}    
    \end{subequations}

    \item With the fixed association strategy, the SINR of $T_0$ without an assisting RIS can be given by
    \begin{equation}
        \gamma_0^{*} = \frac{S_0^{*}}{I + \gamma_t^{-1}},
    \end{equation}
    where
    \begin{equation}
        S_0^{*} = \eta_{g, 0} |g_0|^2.
        \label{s0*}
    \end{equation}

    \item With the nearest association strategy, the SINR of $U_0$ is given by\footnote{$\gamma_0$ is a function of the distance between $T_i$ and $U_0$, $i \in \Lambda_t$, i.e., $d_{g, i}$. With the nearest association strategy, $U_0$ is associated with the TX with the minimum $d_{g, i}$.}
    \begin{equation}
         \gamma_0 = \left\{
            \begin{split}
                &\quad \gamma_0^{+},~~ {\rm w.p.}~~~p, \\[5pt]
                &\quad \gamma_0^{*},~~~ {\rm w.p.}~~~1 - p,
            \end{split}
            \right. 
    \end{equation}
\end{enumerate}

\subsection{Performance Metrics}
In this paper, we will analyze the coverage probability and average achievable rate of the typical user $U_0$.

\subsubsection{Coverage Probability}
Coverage probability is defined as the probability that the received SINR is larger than a threshold, i.e.,
\begin{equation}
    P_c(\bar{\gamma}) = {\mathbb{P}}\left(\gamma > \bar{\gamma}\right),
\end{equation}
where $\gamma \in \{\gamma_0^{+}, \gamma_0^{*}, \gamma_0\}$; $\Bar{\gamma}$ is the coverage SINR threshold.

\subsubsection{Average Achievable Rate} The average achievable rate of the typical user is defined as
\begin{equation}
    R = \int_{0}^{\infty} \log_2\left(1 + x\right) f_{\gamma}(x) {\rmd} x,
    \label{aarr}
\end{equation}
where $f_{\gamma}(x)$ is the PDF of $\gamma$.

\section{Signal Power Analysis}
In this section, we characterize the signal power distributions with and without RIS reflect beamforming, which are denoted as ${S}_0^{+}$ and ${S}_0^{*}$, respectively.

\subsection{Distribution with RIS-Assisted Transmission}
The received signal power with RIS-assisted transmission $S_0^{+}$ can be rewritten as
\begin{equation}
    S_0^{+} = \left(\sqrt{\eta_{g, 0}} |g_0| + \sqrt{\eta_{h, 0}} S_r\right)^2,
    \label{s0+_d}
\end{equation}
where
\begin{equation}
    S_r = \sum\limits_{n = 1}^{N} |h_{0, n}| |r_{0, n}|,
\end{equation}
which is the sum of the products of Nakagami-$m$ random variables. The exact distribution of $S_r$ cannot be obtained in closed form. Hence, we propose a moment matching method based on gamma distribution\footnote{The gamma distribution is often used to fit some complicated SNR distributions, since The gamma distribution gives freedom of adjusting two parameters, and the expectation and variance of the gamma distribution have concise forms, making parameters computationally tractable  \cite{6059452}.} to fit the distribution of $S_r$. 

Since $|h_{0, n}|$ and $|r_{0, n}|$, $n \in \{1,..., N\}$ are independent and identically distributed (i.i.d.) Nakagami-$m$ random variables, the mean and second order raw moment of $S_r$ are derived as:
\begin{subequations}
\begin{align}
    &\mathbb{E}\left[S_r\right] = \sum\limits_{n = 1}^{N} \mathbb{E}\left[|h_{0, n}|\right] \mathbb{E}\left[|r_{0, n}|\right] \nonumber\\
    &= N \frac{\Gamma\left(m_h + \frac{1}{2}\right)}{\Gamma(m_h)}  \frac{\Gamma\left(m_r + \frac{1}{2}\right)}{\Gamma(m_r)} \left(\frac{1}{m_h m_r}\right)^{\frac{1}{2}}, \label{e_s_r}\\
    &\mathbb{E}\left[S_r^2\right] = \mathbb{E}\left[\sum\limits_{n=1}^{N} \left|h_{0, n}\right|^2 \left|r_{0, n}\right|^2 \right] \nonumber\\[3pt]
    &\quad+ 2\, \mathbb{E}\left[\sum\limits_{n=1}^{N-1}\sum\limits_{i=n+1}^{N} \left|h_{0, n}\right|\left|r_{0, n}\right| \left|h_{0, i}\right|\left|r_{0, i}\right|\right] \nonumber\\[3pt]
    &= N + N (N - 1) \times \nonumber\\
    &\left(\frac{\Gamma\left(m_h + \frac{1}{2}\right)}{\Gamma(m_h)}  \frac{\Gamma\left(m_r + \frac{1}{2}\right)}{\Gamma(m_r)}\right)^2 \frac{1}{m_h m_r}. \label{e_s_r1}
\end{align}    
\end{subequations}
Hence, $S_r$ can be approximated by a gamma random variable with shape and scale parameters:
\begin{subequations}
\begin{align}
    \kappa_r &= \frac{\mathbb{E}^2[S_r]}{\mathbb{E}[S_r^2] - \mathbb{E}^2[S_r]},\\
    \omega_r &= \frac{\mathbb{E}[S_r^2] - \mathbb{E}^2[S_r]}{\mathbb{E}[S_r]}.
\end{align}    
\end{subequations}


The moment matching approach can be applied to fit the distribution of ${S}_0^{+}$. Then
the mean and second order moment of ${S}_0^{+}$ can be derived as
\begin{subequations}
\begin{align}
    \mathbb{E}[S_0^{+}] &= \eta_{g,0} \chi_1, \label{e-s-0+}\\
    \mathbb{E}\left[\left(S_0^{+}\right)^2\right] &= \eta_{g,0}^2 \chi_2, \label{e-s-0+2}
\end{align}    
\end{subequations}
where
\begin{subequations}
\begin{align}
    \beta =& \sqrt{\eta_{h, 0}/\eta_{g, 0}},\\
    \chi_1 =& \mu_{2}^g + 2 \beta \mu_{1}^g\,\mu_{1}^{\varsigma} + \beta^2 \mu_{2}^{\varsigma}, \\
    \chi_2 =& \mu_{4}^g + 4 \beta \mu_{3}^g \mu_{1}^{\varsigma} + 6 \beta^2 \mu_{2}^g \mu_{2}^{\varsigma} \nonumber\\
    &+ 4 \beta^3 \mu_{1}^g \mu_{3}^{\varsigma} + \beta^4 \mu_{4}^{\varsigma}.
\end{align}    
\end{subequations}
$\mu_{q}^{\epsilon}$, $\epsilon \in \{g, \varsigma\}$, $q \in \{1, 2, 3, 4\}$ denotes the $q$th order raw moment of $|g_0|$ and $S_r$, respectively. Since $|g_0| \sim \mathcal{R}(\sqrt{2}/2)$ and $S_r$ has been fitted by a gamma random variable, $\mu_{q}^{g}$ and $\mu_{q}^{s}$ are given by
\begin{subequations}
\begin{align}
    \mu_{q}^{g} &= \Gamma\left(1 + \frac{1}{2}q\right), \label{mu-g-q}\\
    \mu_{q}^{{\varsigma}} &= \omega_r^q\,\frac{\Gamma(q + \kappa_r)}{\Gamma(\kappa_r)}. \label{mu-s-q}
\end{align}    
\end{subequations}
Therefore, $S_0^{+}$ can be approximated by a gamma random variable with shape and scale parameters as
\begin{subequations}
\begin{align}
    \kappa_s &= \frac{\chi_1^2}{\chi_2 - \chi_1^2}, \label{kap-s}\\
    \omega_s &= \eta_{g,0} \bar{\chi} = \eta_{g,0} \frac{\chi_2 - \chi_1^2}{\chi_1}. \label{ome-s}
\end{align}    
\end{subequations}
Furthermore, the approximate probability density function (PDF) and cumulative distribution function (CDF) of ${S}_0^{+}$ are given by
\begin{subequations}
\begin{align}
    f_{S_0^{+}}(x) &= \frac{x^{\kappa_s - 1}\,\exp{\left(- x/\omega_s\right)}}{\Gamma (\kappa_s)\,\omega_s^{\kappa_s}}, \label{fs0+}\\
    F_{S_0^{+}}(x) &= \frac{\gamma\left(\kappa_s, x/\omega_s\right)}{\Gamma(\kappa_s)}.\label{Fs0+}
\end{align}    
\end{subequations}

Now, we investigate the channel hardening effect in the proposed system for a large number of elements. The coefficient of variation, i.e., the ratio of the standard deviation to the mean, is summarized in the following corollary.
\begin{corollary}
For a large number of RIS elements, the coefficient of variation of ${S}_0^{+}$ satisfies the asymptotic property:
\begin{equation}
    \nu_s = O\left(\frac{1}{\sqrt{N}}\right),~~~ N \to \infty.
\end{equation}
\label{coro-cv}
\end{corollary}
\begin{IEEEproof}
See Appendix A.
\end{IEEEproof}
\begin{remark}
As introduced in \cite{lyu2021hybrid}, the channel hardening effect refers to the fact that the variation of the signal power around the mean power decreases with the increase of the number of elements. Since the coefficient of variation shows the variability extent of a random variable in relation to its mean value, it can be learned from Corollary~\ref{coro-cv} that increasing the number of elements will lead to less variability of ${S}_0^{+}$ around its mean value. Hence, the channel hardening effect appears. Due to the channel hardening effect, the variation of the channel is averaged out and the overhead of frequent channel estimation is reduced.
\end{remark}

\subsection{Distribution without RIS Beamforming}
Since $|g_0| \sim \mathcal{R}(\sqrt{2}/2)$, $S_0^{*}$ follows the exponential distribution with the distribution parameter $\zeta_{g} = 1/\eta_{g,0}$. The PDF and CDF of $S_0^{*}$ are respectively given by
\begin{subequations}
\begin{align}
    f_{S_0^{*}}(x) = \frac{1}{\eta_{g,0}} e^{-\frac{x}{\eta_{g,0}}},\\
    F_{S_0^{*}}(x) = 1 - e^{-\frac{x}{\eta_{g,0}}}.
\end{align}    
\end{subequations}

\section{Interference Power Analysis}
In this section, the interference power $I$ is analyzed. Referring to \eqref{i_power}, \eqref{i_1kpower} and \eqref{i_2kpower}, we first characterize the distributions of $I_{1, k}$ and $I_{2,k}$, respectively. Then, the Laplace transform of $I$ is computed which is used in analyzing the coverage probability and the average achievable rate of the typical user.

\subsection{Interference Power Distribution Analysis}
\subsubsection{Interference from TX with Assisting RIS}
The interference power from a TX with an assisting RIS $I_{1,k}$ is rewritten by
\begin{equation}
    I_{1,k} = \left|\sqrt{\eta_{g, k}} g_k + \sqrt{\eta_{h, k}} I_{r, k} \right|^2,
\end{equation}
where
\begin{equation}
    I_{r, k} = \sum\limits_{n = 1}^{N} |h_{k, n}| |r_{k, n}| e^{j \theta_{k,n}}.
\end{equation}
Then, the distribution of $I_{r,k}$ is analyzed. Since $\theta_{k,n}$ is uniformly distributed over $[-\pi, \pi)$, the expectation and variance of $|h_{k, n}| |r_{k, n}| \cos(\theta_{k, n})$ and $|h_{k,n}| |r_{k, n}| \sin(\theta_{k, n})$ can be obtained as $\mathbb{E}\left[|h_{k, n}| |r_{k, n}| \cos(\theta_{k, n})\right] = \mathbb{E}\left[|h_{k, n}| |r_{k, n}| \sin(\theta_{k, n})\right] = 0$ and ${\rm Var}\left[|h_{k, n}| |r_{k, n}| \cos(\theta_{k, n})\right] = {\rm Var}\left[|h_{k, n}| |r_{k, n}| \sin(\theta_{k, n})\right] = \frac{1}{2}$.
Further, it can be proved that $|h_{k, n}| |r_{k, n}| \cos(\varphi_{k, n})$ and $|h_{k, n}| |r_{k, n}| \sin(\varphi_{k, n})$ are uncorrelated with each other. 
Using the central limit theorem, for a large number of elements $N$, $I_{r, k}$ follows the circularly symmetric complex Gaussian (CSCG) distribution 
\begin{equation}
    I_{r, k} \sim \mathcal{CN}\left(0, N\right).
\end{equation}
Since $g_k$ follows the CSCG distribution, i.e., $g_k \sim \mathcal{CN}(0, 1)$, it can be learned that
\begin{align}
    \sqrt{\eta_{g, k}} g_k + \sqrt{\eta_{h, k}} I_{r, k} \sim \mathcal{CN}\left(0, \eta_{g, k} + N \eta_{h, k}\right).
\end{align}
Thus, $I_{1, k}$ follows the exponential distribution with the distribution parameter 
\begin{equation}
    \zeta_{k} = \frac{1}{\eta_{g, k} + N \eta_{h, k}}.
\end{equation}
\begin{remark}
Different from the moment matching method based on the gamma distribution used in the analysis of the signal power distribution, the CLT method is used in the analysis of the interference power. There are several reasons for this change. First, due to the RIS-assisted transmission, $S_r$ is the sum of positive random variables. Approximating $S_r$ with the gamma distribution is more reasonable because a gamma random variable is also positive. However, the support of a Gaussian random variable is the set of reals. Hence, using the CLT to analyze $S_r$ will lead to significant errors especially in the lower tail of the distribution. This point is also illustrated in \cite{b9}. In contrast to $S_r$, the phase of each term in $I_{r,k}$ is uniformly distributed over $[-\pi, \pi)$. Therefore, the real and imaginary parts of $I_{r,k}$ are both supported on the real line, and using the CLT to approximate $I_{r, k}$ as a CSCG random variable is more accurate. 
\end{remark}

\begin{remark}
With the CLT, the PDF and CDF of $I_{1,k}$ is unrelated to the scale parameter of the Nakagami-$m$ distribution.
\label{rme3}
\end{remark}

\subsubsection{Interference from TX without Assisting RIS} 
As defined in \eqref{i_1kpower}, Since $|g_k| \sim \mathcal{R}(\sqrt{2}/2)$, $I_{2,k}$ follows the exponential distribution with the distribution parameter $\zeta_{g, k} = 1/\eta_{g,k}$. 
The PDF and CDF of $I_{2, k}$ are given by
\begin{subequations}
\begin{align}
    f_{I_{2, k}}(x) = \frac{1}{\eta_{g,k}} e^{-\frac{x}{\eta_{g,k}}},\\
    F_{I_{2, k}}(x) = 1 - e^{-\frac{x}{\eta_{g,k}}}.
\end{align}    
\end{subequations}

\subsection{Laplace Transform of the Interference Power}
The Laplace transform of the interference power is useful in the analysis of the coverage probability in the next section. Hence, the Laplace transform of the interference power with the fixed and nearest association strategies can be obtained as follows.

\subsubsection{Fixed Association Strategy}
If $U_0$ is associated with a TX at a fixed location, i.e, $d_{g, 0}$ is a fixed value, the Laplace transform of the interference power is summarized in the following proposition.
\begin{proposition}
    The Laplace transform of the interference power $I$ with the fixed association strategy is
    \begin{equation}
         \mathcal{L}_{I, f}(s) = E_{f, 1}(s)\,E_{f, 2}(s),
    \end{equation}
    where
    \begin{subequations}
    \begin{align}
        E_{f, 1}(s) &= \exp\left(-\frac{2 \pi^2 \lambda_t p \csc\left(\frac{2 \pi}{\alpha}\right) (e_1 s)^{\frac{2}{\alpha}}}{\alpha}\right),\\
        E_{f, 2}(s) &= \exp\left(-\frac{2 \pi^2 \lambda_t (1-p) \csc\left(\frac{2 \pi }{\alpha}\right) (C_d s)^{\frac{2}{\alpha}}}{\alpha}\right),\\
        e_1 &= \left(C_d + N C_r d_{0}^{-\alpha}\right).
    \end{align}    
    \end{subequations}
    \label{propp_fix}
\end{proposition}
\begin{IEEEproof}
See Appendix B.
\end{IEEEproof}


\subsubsection{Nearest Association Strategy}
If $U_0$ is associated with the nearest TX, the Laplace transform of the interference power is given in the following proposition.

\begin{proposition}
    The Laplace transform of the interference power $I$ with nearest association strategy is
    \begin{equation}
        \mathcal{L}_{I, n}(s) = E_{n, 1}(s)\,E_{n, 2}(s),
    \end{equation}
    where $E_{n,1}(s)$ and $E_{n,2}(s)$ are given in \eqref{nearest_e1} and \eqref{nearest_e2} at the top of next page, respectively.
    \begin{figure*}
    \begin{subequations}
        \begin{align}
            E_{n,1}(s) &= \exp\left(\pi \lambda_t p d_{g,0}^2 \left(1  - {_{2}F_1}\left(1, - \frac{2}{\alpha}; 1 - \frac{2}{\alpha}; - e_1 d_{g,0}^{-\alpha} s\right)\right) \right), \label{nearest_e1}\\
            E_{n,2}(s) &= \exp\left(\pi \lambda_t (1-p) d_{g,0}^2  \left(1 - {_{2}F_1}\left(1, - \frac{2}{\alpha}; 1 - \frac{2}{\alpha}; - C_d d_{g,0}^{-\alpha} s\right) \right)\right). \label{nearest_e2}
        \end{align}
    \end{subequations}
    \hrule
    \end{figure*}
    $d_{g,0}$ is the distance between $U_0$ and the associated TX that is closest to $U_0$.
    \label{propne}
\end{proposition}
\begin{IEEEproof}
   The proof is similar to the Proposition~\ref{propp_fix} by changing lower limit of the integration to $d_{g,0}$, as given in \eqref{int_lap_near} at the top of next page.
   \begin{figure*}
   \begin{equation}
       \mathcal{L}_{I, n}(s) = \exp\left(2 \pi \lambda_t \int_{d_{g,0}}^{\infty}\left(\frac{p}{1 + s e_1 r^{-\alpha}} + \frac{1 - p}{1 + s C_d r^{-\alpha}} - 1\right)r \rmd r\right) = E_{n, 1}(s)\, E_{n, 2}(s). \label{int_lap_near}
   \end{equation}
   \hrule
   \end{figure*}
\end{IEEEproof}


\section{Performance Analysis}
In this section, we analyze the coverage probability and the average achievable rate of the typical user with the fixed and nearest association strategies. 
\subsection{Coverage Probability Analysis}
\subsubsection{Fixed Association with Assisting RIS}
The main result is concluded in the following proposition.
\begin{proposition}
    If $U_0$ is associated with a TX at a fixed location and the TX has an assisting RIS, the coverage probability of $U_0$ is
    \begin{equation}
        P_{1}(\bar{\gamma}) = \sum\limits_{i = 0}^{\hat{\kappa}_s - 1}\frac{(-1)^i}{i!} \left[\frac{\partial^i}{\partial s^i} \exp\left(V(s)\right)\right]_{s = 1},
    \end{equation}
    where
    \begin{subequations}
    \begin{align}
        \hat{\kappa}_s &= \round{\kappa_s},\\
        V(s) &= -\frac{s \Bar{\gamma}}{\omega_s} \gamma_t^{-1} - \frac{2 \pi^2 \lambda_t p \csc\left(\frac{2 \pi}{\alpha}\right) \left( \frac{e_1 \bar{\gamma}}{\omega_s}s\right)^{\frac{2}{\alpha}}}{\alpha} \nonumber\\
        &~~~ -\frac{2 \pi^2 \lambda_t (1-p) \csc\left(\frac{2 \pi }{\alpha}\right) \left( \frac{C_d \bar{\gamma}}{\omega_s} s\right)^{\frac{2}{\alpha}}}{\alpha}.
    \end{align}    
    \end{subequations}
    \label{propp1}
\end{proposition}
\begin{IEEEproof}
See Appendix C.
\end{IEEEproof}
\begin{remark}
We take another step to round $\kappa_s$ to its nearest positive integer. We will study errors introduced by the rounding operation through simulations.
\end{remark}

\subsubsection{Fixed Association without Assisting RIS} If the associated TX does not have an assisting RIS, the coverage probability is derived in the following proposition.
\begin{proposition}
    If $U_0$ is associated with a TX at a fixed location and the TX does not have an assisting RIS, the coverage probability of $U_0$ is
    \begin{align}
        P_2(\bar{\gamma}) =& \exp\left(-\frac{\bar{\gamma}}{\eta_{g,0}} \gamma_t^{-1} - \frac{2 \pi^2 \lambda_t p \csc\left(\frac{2 \pi}{\alpha}\right) \left( \frac{e_1 \bar{\gamma}}{\eta_{g,0}}\right)^{\frac{2}{\alpha}}}{\alpha} \right. \nonumber\\
        &\quad\quad\quad \left.-\frac{2 \pi^2 \lambda_t (1-p) \csc\left(\frac{2 \pi }{\alpha}\right) \left( \frac{C_d \bar{\gamma}}{\eta_{g,0}} \right)^{\frac{2}{\alpha}}}{\alpha}\right). \label{fixed/cov}
    \end{align}
    \label{prop1-p}
\end{proposition}
\begin{IEEEproof}
    See Appendix D.
\end{IEEEproof}
\begin{remark}
As can be seen from \eqref{fixed/cov}, increasing the density of TXs will lead to lower coverage probability due to the increased interference power.
\end{remark}

\subsubsection{Nearest Association Strategy}
If $U_0$ is associated with the nearest TX, the PDF of $d_{g,0}$ is given by
\begin{equation}
    f_{d_{g,0}}(x) = 2 \pi \lambda_t x e^{-\lambda_t \pi x^2}.
\end{equation}
Then, the coverage probability of $U_0$ is given in the following proposition
\begin{proposition}
    With the nearest association strategy, the coverage probability of $U_0$ is given in an integral form as
    \begin{align}
        P_{c, n}(\bar{\gamma}) &= p \int_{0}^{\infty} q^{+}(r) f_{d_{g,0}}(r) \rmd r \nonumber\\
        &~~~ + (1 - p) \int_{0}^{\infty} q^{*}(r) f_{d_{g,0}}(r) \rmd r,
    \end{align}
    where
    \begin{subequations}
    \begin{align}
        q^{+}(r) &= \sum\limits_{i = 0}^{\hat{\kappa}_s - 1}\frac{(-1)^i}{i!} \left[\frac{\partial^i}{\partial s^i} \exp\left(W(s, r)\right)\right]_{s = 1}, \\
        W(s, r) &=
        -\frac{\Bar{\gamma} \gamma_t^{-1}}{C_d \bar{\chi}}  r^{\alpha} s + \pi \lambda_t r^2 \\
        &~~~ - \pi \lambda_t p r^2 {_{2}F_1}\left(1, - \frac{2}{\alpha}; 1 - \frac{2}{\alpha}; - \frac{e_1 \bar{\gamma}}{C_d \bar{\chi}} s\right) \nonumber\\
        &~~~ - \pi \lambda_t (1 - p) r^2 {_{2}F_1}\left(1, - \frac{2}{\alpha}; 1 - \frac{2}{\alpha}; - \frac{  \bar{\gamma}}{\bar{\chi}} s\right),\nonumber\\
        q^{*}(r) &= \exp\bigg{(}
        -\frac{\bar{\gamma}  \gamma_t^{-1}}{C_d} r^{\alpha} + \pi \lambda_t r^2 \\
        &~~~\quad - \pi \lambda_t p r^2 {_{2}F_1}\left(1, - \frac{2}{\alpha}; 1 - \frac{2}{\alpha}; - \frac{e_1  \bar{\gamma}}{C_d}\right) \nonumber\\
        &~~~\quad - \pi \lambda_t (1 - p) r^2 {_{2}F_1}\left(1, - \frac{2}{\alpha}; 1 - \frac{2}{\alpha}; - \bar{\gamma}\right)
        \bigg{)}.\nonumber
    \end{align}    
    \end{subequations}
    \label{cov_ner1}
\end{proposition}
\begin{IEEEproof}
    See Appendix E.
\end{IEEEproof}

For $\alpha = 4$, the above expression in Proposition~\ref{cov_ner1} can be further simplified, which is summarized in the following Corollary.
\begin{corollary}
    For $\alpha = 4$, the coverage probability of $U_0$ with the nearest association strategy is given in closed form as
    \begin{align}
        P_{c, n}(\bar{\gamma}) &= \frac{\pi \lambda_t p}{2}  \sum\limits_{i = 0}^{\hat{\kappa}_s - 1}\frac{(-1)^i}{i!} \left[\frac{\partial^i}{\partial s^i} X^{+}(s) \right]_{s = 1} \nonumber\\[5pt]
        &~~~ + \frac{\pi \lambda_t (1 - p)}{2} X^{*},
    \end{align}
    where
\begin{subequations}
    \begin{align}
        X^{+}(s) &= \frac{\sqrt{\pi } \exp\left(\frac{{X_{2}^2}(s)}{4 X_1(s)}\right) {\rm erfc}\left(\frac{X_2(s)}{2 \sqrt{X_1(s)}}\right)}{\sqrt{X_1(s)}},\\
        X_1(s) &= \frac{\Bar{\gamma} \gamma_t^{-1}}{C_d \bar{\chi}} s,\\
        X_2(s) &= \pi \lambda_t p\, {_{2}F_1}\left(1, - \frac{1}{2}; \frac{1}{2}; - \frac{e_1 \bar{\gamma}}{C_d \bar{\chi}} s\right) \\
        &~~~ + \pi \lambda_t (1 - p) {_{2}F_1}\left(1, - \frac{1}{2}; \frac{1}{2}; - \frac{  \bar{\gamma}}{\bar{\chi}} s\right),\nonumber\\
        X^{*} &= \frac{\sqrt{\pi } \exp\left(\frac{{X_{4}^2}}{4 X_3}\right) {\rm erfc}\left(\frac{X_4}{2 \sqrt{X_3}}\right)}{\sqrt{X_3}},\\
        X_3 &= \frac{\bar{\gamma}  \gamma_t^{-1}}{C_d},\\
        X_4 &=  \pi \lambda_t p\, {_{2}F_1}\left(1, - \frac{1}{2}; \frac{1}{2}; - \frac{e_1  \bar{\gamma}}{C_d}\right) \nonumber\\
        &~~~+ \pi \lambda_t (1 - p) {_{2}F_1}\left(1, - \frac{1}{2}; \frac{1}{2}; - \bar{\gamma}\right).
    \end{align}
\end{subequations}
\label{coro2_alpha4_cov}
\end{corollary}
\begin{IEEEproof}
    It can be proved by using the interchangeable property of differentiation and integration.
\end{IEEEproof}

Then, we consider the interference-limited case where the noise power is negligible compared to the transmit power $P$, such that $\gamma_t^{-1} \approx 0$. In this case, the closed form outage probability expression can be derived, which is summarized in the following Proposition.
\begin{proposition}
    If $P \gg \sigma_0^2$, the coverage probability of $U_0$ can be expressed as
    \begin{align}
        \Tilde{P}_{c, n}(\bar{\gamma}) &= p  \sum\limits_{i = 0}^{\hat{\kappa}_s - 1}\frac{(-1)^i}{i!} \left[\frac{\partial^i}{\partial s^i} \left( Y_1(s)\right)^{-1}\right]_{s = 1} \nonumber\\[5pt]
        &~~~ + (1-p)  (Y_2)^{-1}
    \end{align}
    where
    \begin{subequations}
    \begin{align}
        Y_1(s) &=  p\, {_{2}F_1}\left(1, - \frac{2}{\alpha}; 1 - \frac{2}{\alpha}; - \frac{e_1 \bar{\gamma}}{C_d \bar{\chi}} s\right) \nonumber\\
        &~~~+  (1 - p)\, {_{2}F_1}\left(1, - \frac{2}{\alpha}; 1 - \frac{2}{\alpha}; - \frac{\bar{\gamma}}{\bar{\chi}} s\right),\\
        Y_2 &=  p\,{_{2}F_1}\left(1, - \frac{2}{\alpha}; 1 - \frac{2}{\alpha}; - \frac{e_1  \bar{\gamma}}{C_d}\right) \nonumber\\
        &~~~+  (1 - p) {_{2}F_1}\left(1, - \frac{2}{\alpha}; 1 - \frac{2}{\alpha}; - \bar{\gamma}\right).
    \end{align}    
    \end{subequations}
    \label{cov_ner2}
\end{proposition}

\begin{IEEEproof}
    As $\gamma_t^{-1} \to 0$, $W(s)$ and $q^{*}(r)$ defined in Proposition~\ref{cov_ner1} can be simplified to \eqref{w_cov_near} and \eqref{q_cov_near} shown at the top of next page.
    \begin{figure*}
    \begin{subequations}
    \begin{align}
        W(s, r) &= \pi \lambda_t r^2  - \pi \lambda_t p r^2 {_{2}F_1}\left(1, - \frac{2}{\alpha}; 1 - \frac{2}{\alpha}; - \frac{e_1 \bar{\gamma}}{C_d \bar{\chi}} s\right)  - \pi \lambda_t (1 - p) r^2 {_{2}F_1}\left(1, - \frac{2}{\alpha}; 1 - \frac{2}{\alpha}; - \frac{  \bar{\gamma}}{\bar{\chi}} s\right), \label{w_cov_near}\\
        q^{*}(r) &= \exp\left(
         \pi \lambda_t r^2  - \pi \lambda_t p r^2 {_{2}F_1}\left(1, - \frac{2}{\alpha}; 1 - \frac{2}{\alpha}; - \frac{e_1  \bar{\gamma}}{C_d}\right)  - \pi \lambda_t (1 - p) r^2 {_{2}F_1}\left(1, - \frac{2}{\alpha}; 1 - \frac{2}{\alpha}; - \bar{\gamma}\right)\right). \label{q_cov_near}
    \end{align}    
    \end{subequations}
    \hrule
    \end{figure*}
    Recalling the definition of $P_{c,n}(\bar{\gamma})$ in Proposition~\ref{cov_ner1} and using the interchangeable property of integration and differentiation, we can obtain the result in Proposition~\ref{cov_ner2}.
\end{IEEEproof}

\begin{remark}
It turns out the coverage probability of $U_0$ does not depend on $\lambda_t$ in the interference-limited scenario. With the nearest association strategy, increasing the density of TXs will help $U_0$ connect with a closer TX. On the other hand, the larger density of TXs will lead to enhanced interference. Proposition~\ref{cov_ner2} shows that these two effects can perfectly offset each other. This point will be further illustrated by numerical results in Sec.VI.
\label{rem_lambda}
\end{remark}


\subsection{Average Achievable Rate Analysis}
The average achievable rate of the typical user can be derived from the coverage probability expression. A lemma regarding the relationship between the average achievable rate and the coverage probability is first provided.

\begin{lemma}
    If the coverage probability of the typical user is $P_c(\bar{\gamma})$, where $\bar{\gamma}$ is the coverage SINR threshold, the average achievable rate of the typical user can be written as
    \begin{align}
        R = \frac{1}{\ln(2)} \int_{0}^{\infty} \frac{P_c(x)}{1 + x} \rmd x.
    \end{align}
    \label{lemmcr}
\end{lemma}
\begin{IEEEproof}
    See Appendix F.
\end{IEEEproof}

With the help of Lemma \ref{lemmcr}, the average achievable rates of the typical user in different cases can be derived directly.
\begin{corollary}
    For the fixed association strategy, the average achievable rate of $U_0$ with and without an assisting RIS can be respectively given as
    \begin{subequations}
    \begin{align}
        R_1 = \frac{1}{\ln(2)} \int_{0}^{\infty} \frac{P_1(x)}{1 + x} \rmd x,\\
        R_2 = \frac{1}{\ln(2)} \int_{0}^{\infty} \frac{P_2(x)}{1 + x} \rmd x,
    \end{align}
    \end{subequations}
    where $P_1(x)$ and $P_2(x)$ are defined in Proposition~\ref{propp1} and Proposition~\ref{prop1-p}, respectively.
\end{corollary}

If $\alpha = 4$, $R_1$ and $R_2$ in the interference-limited scenario can be further transformed in the following Corollary.
\begin{corollary}
    In an interference-limited scenario with $\alpha = 4$, $R_1$ and $R_2$ are given by
    \begin{subequations}
        \begin{align}
            \Tilde{R}_1 &= \frac{1}{\ln(2)} \sum\limits_{i = 0}^{\hat{\kappa}_s - 1}\frac{(-1)^i}{i!} \left[\frac{\partial^i}{\partial s^i} V^{+}(s) \right]_{s = 1} \\
            \Tilde{R}_2 &= \frac{1}{\ln(2)} \left((\pi -2 {\rm Si}(V_2(s))) \sin(V_2(s))\right. \nonumber\\[5pt]
            &~~~ \left.-2 {\rm Ci}(V_2(s)) \cos(V_2(s))\right)
        \end{align}
    \end{subequations}
    where
    \begin{subequations}
        \begin{align}
            V^{+}(s) &= (\pi -2 {\rm Si}(V_1(s))) \sin(V_1(s)) \nonumber\\[5pt]
            &~~~ -2 {\rm Ci}(V_1(s)) \cos(V_1(s)), \\
            V_1(s) &= \frac{\pi^2}{2} \lambda_t \left( p  \left( \frac{e_1 }{\omega_s}s\right)^{\frac{1}{2}} + (1-p)  \left( \frac{C_d }{\omega_s} s\right)^{\frac{1}{2}}\right),\\
            V_2(s) &= \frac{\pi^2}{2} \lambda_t \left( p \left( \frac{e_1 }{\eta_{g,0}}\right)^{\frac{1}{2}} + (1-p)  \left( \frac{C_d }{\eta_{g,0}} \right)^{\frac{1}{2}}\right).
        \end{align}
    \end{subequations}
\end{corollary}
\begin{IEEEproof}
    It can be proved by using the interchangeable property of differentiation and integration.
\end{IEEEproof}

\begin{corollary}
    For the nearest association strategy, the average achievable rate of $U_0$ is given by
    \begin{align}
        R_{c, n} = \frac{1}{\ln(2)} \int_{0}^{\infty} \frac{P_{c, n}(\bar{\gamma})}{1 + \bar{\gamma}} \rmd \bar{\gamma},
    \end{align}
    where $P_{c, n}(\bar{\gamma})$ is defined in Proposition~\ref{cov_ner1}.
\end{corollary}
\begin{corollary}
    For the nearest association strategy, the average achievable rate of $U_0$ in the interference-limited scenario is given by
    \begin{align}
        \Tilde{R}_{c, n} = \frac{1}{\ln(2)} \int_{0}^{\infty} \frac{\Tilde{P}_{c, n}(x)}{1 + x} \rmd x,
    \end{align}
    where $\Tilde{P}_{c, n}(x)$ is defined in Proposition~\ref{cov_ner2}.
\end{corollary}

\section{Numerical Results}
In this section, we present numerical results to validate the effectiveness of the above analysis, including the approximate signal power distribution, coverage probability and average achievable rate. We also explore several insights regarding the impacts of different system parameters on the communication performance. 

\subsection{Simulation Setup}
Unless otherwise mentioned, the parameters of the Monte Carlo (MC) simulations are listed as follows. The GPP for the distribution of TXs and RISs is generated in a disk area of radius $5000\,{\rm m}$. The distance between a RIS and its corresponding TX and the path loss exponent are set to be $d_0 = 3 {\rm}$ and $\alpha = 2.5$, respectively. The path loss per unit distance of the direct link and the reflected path are given by $C_d = C_r = -30\,{\rm dB}$. $U_0$ is located at the origin. If the fixed association strategy is applied, the locations of $U_0$'s corresponding TX and RIS are assumed to be fixed with the coordinates $(20, 0)$ and $(20, 3)$, respectively. The AWGN power is $\sigma_0^2 = -70\,{\rm dBm}$.

\subsection{Analysis of Approximate Distributions}
\begin{figure}[!]
    \centerline{\includegraphics[scale=0.6]{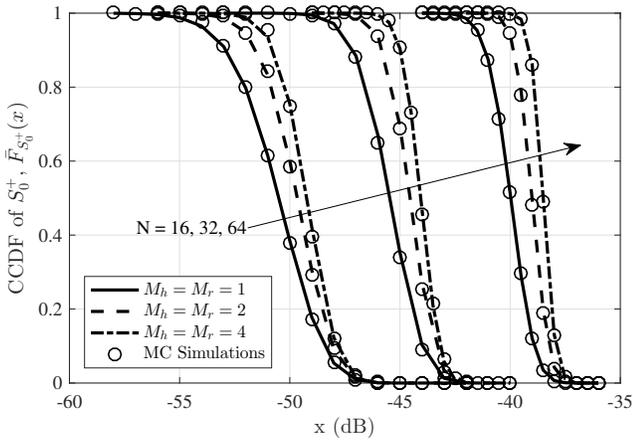}}   
    \caption{Approximate CCDF of $S_0^{+}$ and simulation results.}
    \label{cdf_approx_fig}
\end{figure}

The accuracy of approximating the received signal power $S_0^{+}$ as a gamma random variable is verified here. The shape parameters of the Nakagami-$m$ distribution\footnote{For clarity, we use $m_h = m_r$ in simulations. However, $m_h$ and $m_r$ can be set arbitrarily and independently.} vary in $m_h, m_r \in \{1, 2, 4\}$, and the number of elements varies in $N \in \{16, 32, 64\}$. Fig.~\ref{cdf_approx_fig} illustrates the CCDF of $S_0^{+}$ with different numbers of elements and different shape parameters. The proposed fitting method based on a gamma distribution matches very well with the simulation results. It can be observed that increasing the number of elements can significantly improve the channel power gain. For example, when $m_h = m_r = 1$ and $\bar{F}_{S_0^{+}}(x) = 0.8$, we can see that $x = -52\,{\rm dB}$ and $x = - 41\,{\rm dB}$ for $N = 16$ and $N = 64$, respectively. The channel hardening effect is also illustrated since the CCDF curves become steeper with the increase of the number of elements. 

\begin{figure}[!]
    \centerline{\includegraphics[scale=0.6]{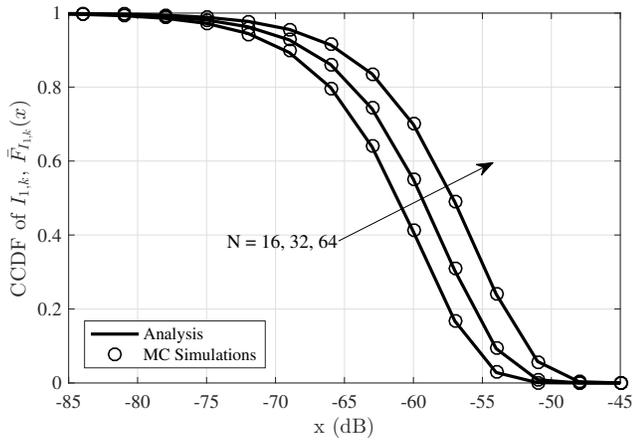}}   
    \caption{Approximate CCDF of $I_{1, k}$ and simulation results.}
    \label{inf_cdf_approx_fig}
\end{figure}

We also investigate the approximation of the interference power from an interfering TX with an assisting RIS, $I_{1, k}$. 
In order to illustrate the performance gain brought by RIS reflect beamforming, we assume the coordinates of the interfering TX and its corresponding RIS are also $(0, 20)$ and $(3, 20)$, respectively. Furthermore, as indicated in Remark~\ref{rme3}, the shape parameter of the Nakagami-$m$ distribution will not affect the interference distribution due to the lack of reflect beamforming. Fig.~\ref{inf_cdf_approx_fig} gives the CCDF of $I_{1, k}$ with different numbers of elements. Compared with the case with reflect beamforming in Fig.~\ref{cdf_approx_fig}, the interference power is much smaller even if the interfering TX without beamforming is at the same distance from $U_0$ as the associated TX. Another important point is that the interference power will not grow as rapidly with the increase of the elements as the case with reflect beamforming. This shows that using RIS with a large number of elements can significantly increase the received SINR in a multi-cell network.


\subsection{Fixed Association Strategy Analysis}
\begin{figure}[!]
    \centerline{\includegraphics[scale=0.6]{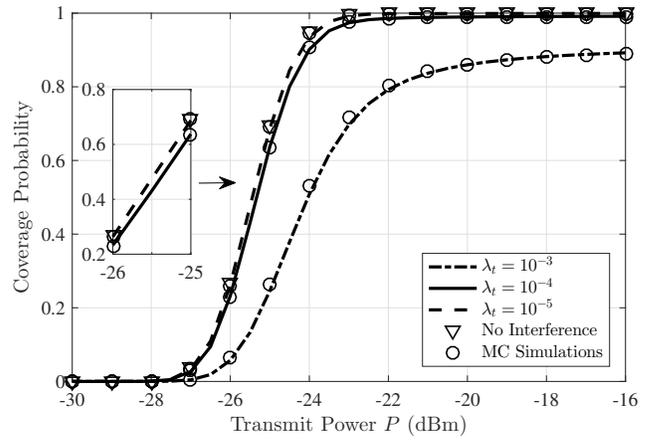}}   
    \caption{Coverage probability versus transmit power with different densities of TXs. Fixed association strategy with assisting RIS.}
    \label{cov_vs_density_fig}
\end{figure}

Fig.~\ref{cov_vs_density_fig} illustrates the coverage probability versus the transmit power with different densities of TXs. $U_0$ is associated with a fixed TX that has an assisting RIS. The number of elements of each RIS is $N = 32$, and the RIS association probability is $p = 0.5$. The Nakagami-$m$ shape parameters are set to be $m_h = m_r = 2$. To illustrate the impact of interfering TXs, the coverage probability curve without interfering TXs, labelled as `No Interference', is also given. We can see that, even though several approximations are applied to obtain the closed form coverage probability in Proposition~\ref{propp1}, the proposed analytical result is still accurate for a large density of TXs, e.g., $\lambda_t = 10^{-3}$. Interfering TXs sharing the same resource block will always deteriorate the system performance, but the influence is only significant when the density of TXs is large. For example, when the transmit power is $P = -24 {\rm dBm}$, the coverage probability is $0.53$ and $0.93$ for $\lambda_t = 10^{-3}$ and $\lambda_t = 10^{-4}$, respectively. However, the impact of interfering TXs is negligible when $\lambda_t < 10^{-4}$ as the curves are close to the ideal case without interference. 

\begin{figure}[!]
    \centerline{\includegraphics[scale=0.6]{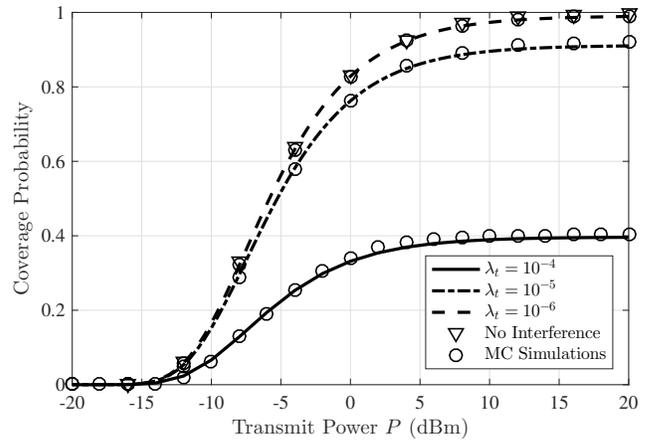}}   
    \caption{Coverage probability versus transmit power with different densities of TXs. Fixed association strategy without assisting RIS.}
    \label{cov_vs_density_woRIS_fig}
\end{figure}

Fig.~\ref{cov_vs_density_woRIS_fig} shows the coverage probability versus the transmit power when $U_0$ is associated with a fixed TX without an assisting RIS. The performance degradation due to interfering TXs is negligible if $\lambda_t < 10^{-6}$. Compared with Fig.~\ref{cov_vs_density_fig}, we can learn that associating $U_0$ with a TX that has an assisting RIS can significantly improve the performance. For example, if $\lambda_t = 10^{-5}$, achieving the coverage probability of $0.9$ requires the transmit power of $P = -24\,{\rm dBm}$ and $10\,{\rm dBm}$ for the cases with and without assisting RIS, respectively.

\begin{figure}[!]
    \centerline{\includegraphics[scale=0.6]{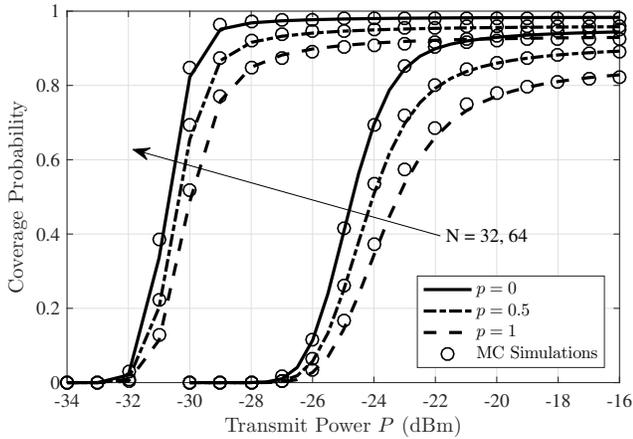}}   
    \caption{Coverage probability versus transmit power with different RIS association probabilities. Fixed association strategy with assisting RIS.}
    \label{cov_vs_probability_fig}
\end{figure}

The impact of the RIS association probability $p$ is explored in Fig.~\ref{cov_vs_probability_fig}. $U_0$ is associated with a fixed TX with an assisting RIS. The RIS association probability and the number of elements vary in $p \in \{0, 0.5, 1\}$ and $N \in \{32, 64\}$, respectively. As expected, increasing $p$ will lead to lower coverage probability if the fixed associations strategy is applied since the interference power scattered by RISs is enhanced.

\begin{figure}[!]
    \centerline{\includegraphics[scale=0.6]{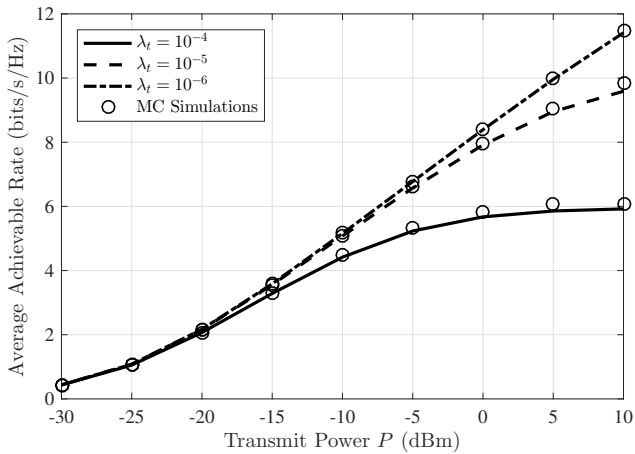}}   
    \caption{Average achievable versus transmit power with different densities of TXs. Fixed association strategy.}
    \label{cap_fixed_density_fig}
\end{figure}

Fig.~\ref{cap_fixed_density_fig} shows the average achievable rate versus the transmit power with different densities of TXs. $U_0$ is connected with a fixed TX that has an assisting RIS. The RIS association probability is $p = 0.5$, and the number of elements per RIS is $N = 32$. The density of TXs varies in $\lambda_t \in \{10^{-6}, 10^{-5}, 10^{-4}\}$. As expected, increasing the transmit power and reducing the density of TXs will lead to better performance of $U_0$.

\subsection{Nearest Association Strategy Analysis}
\begin{figure}[!]
    \centerline{\includegraphics[scale=0.6]{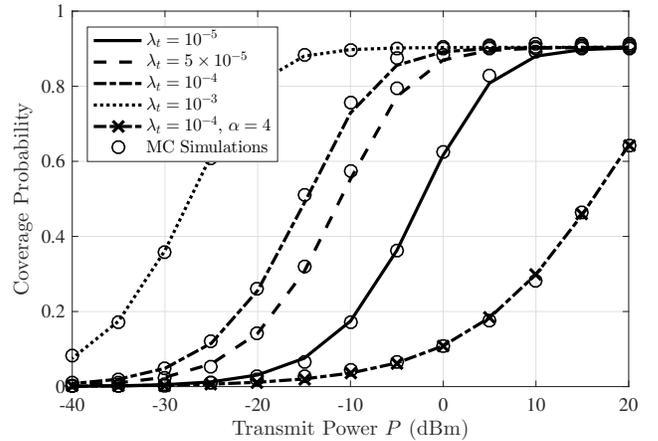}}   
    \caption{Coverage probability versus transmit power with different densities of TXs. Nearest association strategy.}
    \label{cov_vs_nearest_fig}
\end{figure}

\begin{figure}[!]
    \centerline{\includegraphics[scale=0.6]{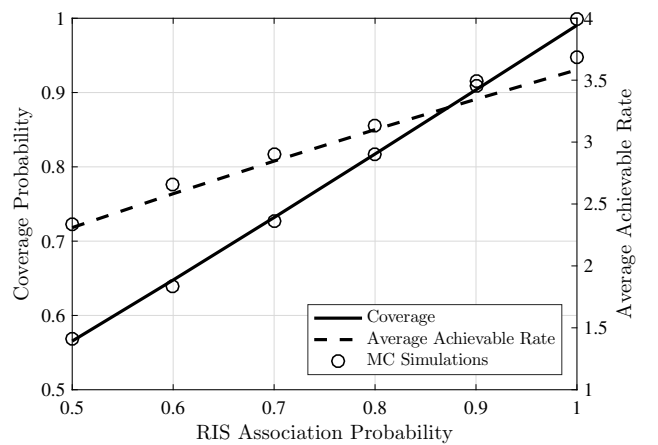}}   
    \caption{Coverage probability and average achievable rate versus RIS association probability. Nearest association strategy.}
    \label{cov_nearest_probability_fig}
\end{figure}

Fig.~\ref{cov_vs_nearest_fig} illustrates the coverage probability against the transmit power when $U_0$ is associated with the nearest TX. The density of TXs varies in $\lambda_t \in \{10^{-5}, 5\times 10^{-5}, 10^{-4}, 10^{-3}\}$, and the RIS association probability is $p = 0.9$. The number of elements per RIS is $N = 32$. To verify the analysis in Corollary \ref{coro2_alpha4_cov}, the curve for $\alpha = 4$ is also given. Contrary to the case of the fixed association strategy, increasing the density of TXs is beneficial for $U_0$ within certain transmit power regime, since $U_0$ is more likely to be connected with a closer TX with the nearest association strategy. Furthermore, we can observe that the coverage probability is independent with the density of TXs when the noise power is negligible, which proves our analysis in Remark~\ref{rem_lambda}.
As a step further, we investigate the impact of the RIS association probability $p$ on the coverage probability and the average achievable rate in the interference-limited scenario, i.e., $\gamma_t \to \infty$, with the nearest association strategy in Fig.~\ref{cov_nearest_probability_fig}. In the current simulation setup, it turns out that increasing the RIS association probability can also improve the system performance since $U_0$ is more likely to be served by a RIS-assisted TX.

\section{Conclusions}
This paper proposed a RIS-assisted multi-cell network with large-scale randomly deployed TXs, RISs and UEs. The locations of TXs and RISs were modeled by a GPP to characterize the spatial correlations. By utilizing the stochastic geometry, we investigated the coverage probability and the average achievable rate of the typical user. The analytical results showed that increasing the density of TXs has no influence on the coverage probability and the average achievable rate in the large transmit SNR regime with the nearest association strategy. The analysis was validated by the numerical results, which also demonstrated the significant performance gain brought by the RIS assisted transmission with passive beamforming, and illustrated that increasing the density of TXs and the RIS association probability was detrimental and beneficial for the system performance with the fixed and nearest association strategies, respectively.

\begin{appendices}
\setcounter{equation}{0}
\renewcommand\theequation{A.\arabic{equation}}
\section{}
Since $S_0^{+}$ is approximated by a gamma random variable with shape and scale parameters $\kappa_s$ and $\omega_s$, we can have
\begin{equation}
    \nu_s^2 = \frac{1}{\kappa_s} = \frac{\chi_2 - \chi_1^2}{\chi_1^2}.
\end{equation}
By using the multinomial theorem, it can be proved that $\chi_2 - \chi_1^2$ and $\chi_1^2$ can be given in polynomials as
\begin{subequations}
    \begin{align}
        \chi_2 - \chi_1^2 &= a_0 + a_1 N + a_2 N^2 + a_3 N^3,\\
        \chi_1^2 &= a'_0 + a'_1 N + a'_2 N^2 + a'_3 N^3 + a'_4 N^4,
    \end{align}
\end{subequations}
where $a_i$, $i \in \{0, 1, 2, 3\}$ and $a'_i$, $i \in \{0, 1, 2, 3, 4\}$ are some real coefficients.
Therefore, we can have
\begin{equation}
    \frac{1}{\kappa_s} \sim \frac{a_3}{a'_4 N},~~~ N \to \infty.
\end{equation}

\setcounter{equation}{0}
\renewcommand\theequation{B.\arabic{equation}}
\section{}
The Laplace transform of the interference power $I$ with the fixed association strategy can be written as
\begin{align}
    &\mathcal{L}_{I, f}(s) = \mathbb{E}\left[e^{-s I}\right] \\
    &= \mathbb{E}\left[\exp\left(-s \left(\sum\limits_{k \in \Lambda_t\setminus \{0\}} I_{1, k} b_k + I_{2, k} (1 - b_k)\right) \right)\right] \nonumber\\
    &\overset{(a)}{=} \mathbb{E}\left[\prod\limits_{k \in \Lambda_t\setminus \{0\}} \left( p \exp\left(-s I_{1, k}\right) + (1-p) \exp\left(-s I_{2, k}\right) \right)\right] \nonumber\\
    &\overset{(b)}{=} \mathbb{E}\left[\prod\limits_{k \in \Lambda_t\setminus \{0\}} \left(\frac{p}{1 +\left(\eta_{g, k} + N \eta_{h, k} \right)s} + \frac{1 - p}{1 + \eta_{g, k} s} \right)\right],\nonumber\\
    &\overset{(c)}{\approx} \mathbb{E}\left[\prod\limits_{k \in \Lambda_t\setminus \{0\}} \left(\frac{p}{1 + e_1 d_{g, k}^{-\alpha} s} + \frac{1 - p}{1 + C_d d_{g, k}^{-\alpha} s} \right)\right] \nonumber\\
    & \overset{(d)}{=} \exp\left(2 \pi \lambda_t  \int_{0}^{\infty}\left(\frac{p}{1 + s e_1 r^{-\alpha}} + \frac{1 - p}{1 + s C_d r^{-\alpha}} - 1\right)r \rmd r\right) \nonumber\\
    &= \exp{\left(2\pi \lambda_t p \int_{0}^{\infty} \left(\frac{1}{1 + s e_1 r^{-\alpha}} - 1\right)r \rmd r \right)}  \nonumber\\
    &~~~\times \exp{\left(2 \pi \lambda_t (1 - p)\int_{0}^{\infty} \left(\frac{1}{1 + s C_d r^{-\alpha}} - 1\right)r \rmd r \right)}, \nonumber
\end{align}
where $(a)$ follows from Lemma 2 in \cite{guo2016gauss}; $(b)$ is obtained by the fact that $I_{1, k}$ and $I_{2, k}$ both follow the exponential distribution.
From the Slivnyak's theorem, conditioning on a TX at a fixed location does not influence the distribution of other clusters; $(c)$ is due to the distance approximation. Considering the spatial correlation between the locations of a TX and its associated RIS where a separate link is required to pass the CSI, $d_{g,j}$ can be significantly larger than $d_0$. A similar assumption is also applied in \cite{lyu2021hybrid}. Besides, $(d)$ holds by using the probability generating functional (PGFL). Thus, the result in Proposition \ref{propp_fix} can be obtained.

\setcounter{equation}{0}
\renewcommand\theequation{C.\arabic{equation}}
\section{}
We have used the moment matching approach to fit the distribution of $S_0^{+}$ to a gamma random variable with shape parameter $\kappa_s$ and scale parameter $\omega_s$ in Sec.III. Here, we round $\kappa_s$ to its nearest integer $\hat{\kappa}_s$. Then, we can have
\begin{align}
    P_1(\bar{\gamma}) &= \mathbb{P}\left(\gamma_0^{+} > \bar{\gamma}\right) \nonumber\\
    &= \mathbb{P}\left(S_0^{+} > \Bar{\gamma} (I + \gamma_t^{-1})\right) \nonumber\\
    &= \mathbb{E}_{I}\left[\bar{F}_{S_0^{+}}(\Bar{\gamma} (I + \gamma_t^{-1}))\right] \nonumber\\
    &= \mathbb{E}_{X}\left[\frac{\Gamma\left(\hat{\kappa}_s, X\right)}{\Gamma(\hat{\kappa}_s)}\right],
\end{align}
where 
\begin{equation}
    X = \frac{\bar{\gamma}}{\omega_s}\left(I + \gamma_t^{-1}\right).
\end{equation}
Since $\hat{\kappa}_s$ is an integer, using the series expansion of the upper incomplete gamma function, we can write
\begin{align}
    P_1(\bar{\gamma}) &= \sum\limits_{i = 0}^{\hat{\kappa}_s - 1}\frac{(-1)^i}{i!} \mathbb{E}_X \left[(-1)^i X^i e^{-X}\right] \nonumber\\
    &= \sum\limits_{i = 0}^{\hat{\kappa}_s - 1}\frac{(-1)^i}{i!} \mathbb{E}_X\left[\left[\frac{\partial^i}{\partial s^i} e^{-s X }\right]_{s = 1}\right] \nonumber\\
    &\overset{(a)}{=} \sum\limits_{i = 0}^{\hat{\kappa}_s - 1}\frac{(-1)^i}{i!} \left[\frac{\partial^i}{\partial s^i} \mathbb{E}_X\left[e^{-s X}\right]\right]_{s = 1} \nonumber\\
    &= \sum\limits_{i = 0}^{\hat{\kappa}_s - 1}\frac{(-1)^i}{i!} \left[\frac{\partial^i}{\partial s^i} \mathcal{L}_{X}(s)\right]_{s = 1} ,
    \label{e_x_ccdf}
\end{align}
where $(a)$ follows by interchanging differentiation and expectation. 

The Laplace transform of $X$ can be derived directly from the Laplace transform of $I$:
\begin{align}
    \mathcal{L}_X(s) &= \mathbb{E}_{I}\left[\exp{\left(-\frac{s \Bar{\gamma}}{\omega_s} (I + \gamma_t^{-1})\right)}\right] \nonumber\\
    &= \exp{\left(-\frac{s \Bar{\gamma}}{\omega_s} \gamma_t^{-1}\right)} \mathcal{L}_{I,f}\left(\frac{s \Bar{\gamma}}{\omega_s}\right).
\end{align}
Using the result in Proposition~\ref{propp_fix}, we arrive at the expression of $P_1$ given in the proposition.

\setcounter{equation}{0}
\renewcommand\theequation{D.\arabic{equation}}
\section{}
Since $S_0^{*}$ is exponentially distributed, we can write
\begin{align}
    P_2 &= \mathbb{P}\left(\gamma_0^{*} > \bar{\gamma}\right) \nonumber\\
    &= \mathbb{P}\left(S_0^{*} > \Bar{\gamma} (I + \gamma_t^{-1})\right) \nonumber\\
    &= \mathbb{E}_{I}\left[\bar{F}_{S_0^{*}}(\Bar{\gamma} (I + \gamma_t^{-1}))\right] \nonumber\\
    &= \mathbb{E}_{I}\left[\exp\left(-\frac{\bar{\gamma}}{\eta_{g,0}} \left(I + \gamma_t^{-1}\right)\right)\right] \nonumber\\
    &= \exp\left(-\frac{\bar{\gamma}}{\eta_{g,0}} \gamma_t^{-1}\right) \mathcal{L}_{I,f}\left(\frac{\bar{\gamma}}{\eta_{g,0}}\right).
\end{align}
Thus, the proof is complete.

\setcounter{equation}{0}
\renewcommand\theequation{E.\arabic{equation}}
\section{}
The coverage probability of $U_0$ can be transformed as
\begin{align}
    P_{c, n}(\bar{\gamma}) &= \mathbb{E}_{d_{g,0}}\left[\mathbb{P}\left(\gamma_0 > \bar{\gamma}\right)\right] \nonumber\\
    &= \mathbb{E}_{d_{g,0}}\left[p \mathbb{P}\left(\gamma_0^+ > \bar{\gamma}\right) + (1 - p)\mathbb{P}\left(\gamma_0^* > \bar{\gamma}\right) \right] \nonumber\\
    &= p \int_{0}^{\infty} \mathbb{P}\left(\gamma_0^+ > \bar{\gamma}\right) f_{d_{g,0}}(r) \rmd r \nonumber\\
    &~~~+ (1 - p) \int_{0}^{\infty} \mathbb{P}\left(\gamma_0^* > \bar{\gamma}\right) f_{d_{g,0}}(r) \rmd r.
\end{align}
The distance approximation used in the proof of Proposition~\ref{propp_fix} such that $d_{g, j} \approx d_{r, j}$, $j\in \Lambda_2^{(P)}$ is applied here. Then, following a similar procedure to Proposition~\ref{propp1} and Proposition~\ref{prop1-p}, using the result in Proposition~\ref{propne}, and recalling the definitions of $\omega_s$ and $\eta_{g,0}$, we can directly derive $\mathbb{P}\left(\gamma_0^+ > \bar{\gamma}\right)$ and $\mathbb{P}\left(\gamma_0^* > \bar{\gamma}\right)$ as $q^{+}(r)$ and $q^{*}(r)$, respectively.

\setcounter{equation}{0}
\renewcommand\theequation{F.\arabic{equation}}
\section{}
As defined in \eqref{aarr}, the average achievable rate of the typical user can be rewritten as
\begin{align}
    R &\overset{(a)}{=} \int_{0}^{\infty} \log_2\left(1 + \bar{\gamma}\right) \frac{\partial (1 - P_c(\bar{\gamma}))}{\partial \bar{\gamma}} {\rmd} \bar{\gamma} \nonumber\\
    &= -\int_{0}^{\infty} \log_2\left(1 + \bar{\gamma}\right) \frac{\partial  P_c(\bar{\gamma})}{\partial \bar{\gamma}} {\rmd} \bar{\gamma} \nonumber\\
    &= -\frac{1}{\ln{(2)}}\int_{0}^{\infty} \int_{0}^{\bar{\gamma}} \frac{1}{1 + x} {\rmd x} \frac{\partial  P_c(\bar{\gamma})}{\partial \bar{\gamma}} {\rmd} \bar{\gamma} \nonumber\\
    &\overset{(b)}{=} -\frac{1}{\ln{(2)}}\int_{0}^{\infty} \frac{1}{1 + x} \int_{x}^{\infty} \frac{\partial  P_c(\bar{\gamma})}{\partial \bar{\gamma}} {\rmd \bar{\gamma}} {\rmd x} \nonumber\\
    &= \frac{1}{\ln{(2)}}\int_{0}^{\infty} \frac{P_c(x)}{1 + x}  {\rmd x},
\end{align}
where $(a)$ is due to the mathematical relationship between the PDF and CDF; $(b)$ is obtained by changing the integration order.

\end{appendices}

\bibliographystyle{IEEEtran}
\bibliography{irs}

\end{document}